\documentclass[aps,prd,twocolumn,showpacs,floatfix]{revtex4}
\usepackage{amsmath}
\usepackage{amsthm}
\usepackage{amsfonts}
\usepackage{amssymb}
\usepackage{graphicx}

\begin{document}
\title{A path-integral representation of the free one-flavour staggered-fermion
determinant}
\author{Francesca Maresca\footnote{Present address: Physics Department,
University of Utah, Salt Lake City, UT 84112, USA} and Mike Peardon}
\affiliation{School of Mathematics, Trinity College, Dublin 2, Ireland.}
\date{\today}

\begin{abstract}
Lattice fermion actions are constructed with path integrals which are
equivalent to the free one-flavour staggered fermion determinant. The
Dirac operators used are local and have an identical spectrum of states to the 
staggered theory. Operators obeying a generalised Ginsparg-Wilson relation are 
developed.
\end{abstract}
\maketitle

\section{Introduction}

A complete study of QCD on the lattice requires the numerical simulation of dynamical 
fermions. These Monte Carlo calculations are extremely computationally costly,
since the effects of quarks must be included by first integrating out the
fermion path integral, and then describing the resulting non-local dynamics of
the fermion determinant. For a review of recent developments, see {\it e.g.} 
\cite{Kennedy:2004ae,Jansen:2003nt}. With present techniques, 
the most cost-effective means of performing these simulations is to use the 
staggered fermion formulation of Kogut and Susskind \cite{Kogut:1974ag}. Recent 
calculations by the MILC collaboration \cite{Davies:2003ik} have demonstrated 
good agreement between experimentally known strong-interaction measurements and 
staggered fermion lattice QCD simulation. 

The formulation as it stands has a serious deficiency for dynamical
simulations. In four dimensions, the staggered fermion determinant describes 
four flavours of fermion, not one. This means that while it is very simple to 
simulate four mass-degenerate fermions with the staggered method, the study of 
one or two flavours must use a fractional power of the fermion determinant.
This raises difficult theoretical problems: what are the fermion fields, and
what is the local action on these fields which reproduces this determinant? 
Without a path-integral representation of the fermion determinant, all the 
standard quantum field theory construction of propagators (which are the 
two-point functions of the underlying quark fields) is poorly defined. If no 
local action exists, an even more severe issue arises, since there is then no 
guarantee that the continuum limit of the lattice simulation is in the same 
universality class as QCD and the link with physics is lost. 

In this paper, we describe a numerical construction of an operator that defines
a lattice quantum field theory equivalent to a single, free staggered fermion.
Most of the construction is performed in two dimensions to ease the
computations, but some suggestive results in four dimensions indicate the
same construction works there too. Note that all the work in this paper is for
the theory of free fermions, and the question of defining the interacting 
theory remains open. 
We do however regard this as a useful starting point for the more difficult 
problem of finding a path integral representation of the staggered fermion 
determinant in the presence of background gauge fields and the construction 
presented does suggest how to proceed further. The paper is organised as 
follows: Sec. \ref{sec:stag_theory} briefly describes the free staggered 
fermion, and Sec. \ref{sec:dirac} describes the numerical construction of the 
local operator.  Sec. \ref{sec:gw} then presents a different numerical 
construction that is seen to obey a modified Ginsparg-Wilson relation. In 
Secs. \ref{sec:discuss} and \ref{sec:conclude} a discussion of our results and 
conclusions is given.

\section{Staggered Fermions \label{sec:stag_theory}}

In this section, a brief overview of the staggered fermion formulation of Kogut
and Susskind \cite{Kogut:1974ag} is presented, emphasising some critical 
properties of the resulting free quark operator. 

The staggered fermion formalism is constructed by first writing a naive
representation of the Dirac operator on the lattice
\begin{equation}
  M_{x,y}^{i,j} = a m \delta_{x,y} \delta^{i,j} + \frac{1}{2}\sum_{\mu} 
\left(\gamma_\mu\right)^{i,j}
\left(
\delta_{x+\hat{\mu},y} - \delta_{x-\hat{\mu},y} 
\right),
\end{equation}
where the Euclidean space indices ($x,y$) and Dirac algebra indices ($i,j$)
have been included explicitly.
This operator has poles not only at zero momentum, but also at the corners of
the Brillouin zone. Counting these poles suggests the field coupled through this
interaction matrix can be thought of as representing $2^d$ flavours of fermions
in $d$ dimensions. 
A local change of variable at every site of the lattice,
$\chi(x) = T(x) \psi(x)$ with 
\begin{equation}
  T(x) = \prod_{\mu=1}^{d} \left(\gamma_\mu\right)^{x_\mu},
\end{equation}
diagonalises the naive operator $M$ over the Dirac algebra. 
The number of flavours is reduced by discarding all but one of the diagonal
components of $\chi$. The operator is 
then regarded as acting on $n_t = 2^d / 2^{d/2} = 2^{d/2}$ flavours of 
fermions. These flavours, all of which appear in a single instance of the 
staggered field, are often termed ``tastes''. The operator on the field is then 
\begin{equation}
  Q_{x,y} = a m \delta_{x,y} + \frac{1}{2} \sum_{\mu} 
\eta_\mu(x)
\left(
\delta_{x+\hat{\mu},y} - \delta_{x-\hat{\mu},y} 
\right).
\end{equation}
where $\eta_\mu(x)$ is the staggered phase, given by 
\begin{equation}
\eta_\mu(x) = (-1)^{\sum_{i=1}^{\mu-1} x_i}.
\end{equation}

A natural decomposition for the operator is to break the lattice into 
hypercubes of side-length $b=2a$ containing $2^d$ sites. A site on the full 
lattice can be labelled with co-ordinates
\begin{equation}
   x_\mu = 2 N_\mu + \rho_\mu,
\end{equation}
where $N_\mu$ are the co-ordinates of sites on the blocked lattice and
$\rho_\mu \in \{ 0,1 \}$. The $2^d$ staggered variables in a hypercube
can be labelled by these hypercubic offset vectors, $\rho$
\begin{equation}
   \chi_\rho(N) = \chi(2 N + \rho).
\end{equation}
Introducing a new spinor field $\psi^{ab}$ on the blocked lattice sites, $N$
\begin{equation}
  \psi^{ab}(N) = \sum_{\rho} [T_\rho]^{ab} \chi_\rho(N),
\end{equation}
gives the staggered fermion action in terms of these 
$2^{d/2}$ tastes of Dirac spinors,
\begin{equation}
  S_{\rm stag} =  b^4 \sum_{N,N'} \bar{\psi}(N) Q(N,N') \psi(N'),
\end{equation}
with 
\begin{widetext}
\begin{equation}
  Q(N,N') = m (I \otimes I) \delta_{N,N'} + \sum_\mu 
      (\gamma_\mu \otimes I) \Delta_\mu(N,N')
   +  \frac{1}{2} b (\gamma_5 \otimes t_\mu t_5) \Box_\mu(N,N'),
\end{equation}
\end{widetext}
where $\Delta_\mu$ and $\Box_\mu$ are the simplest representations of the first
and second derivatives on the blocked lattice, 
\begin{equation}
   \Delta_\mu(N,N') = \frac{1}{2b} (\delta_{N+\mu,N'} - \delta_{N-\mu,N'}),
\end{equation}
\begin{equation}
   \Box_\mu(N,N') = \frac{1}{b^2} (\delta_{N+\mu,N'} + \delta_{N-\mu,N'} 
                        - 2 \delta_{N,N'}).
\end{equation}

This representation makes the Dirac and taste structure of the staggered
operator more apparent. Taking a direct fractional power of a matrix does 
not change its structure, so it seems counterintuitive to expect the matrix
$Q^{1/n_t}$ to be a sensible representation of the one-flavour Dirac operator,
and this operator has been shown to be non-local \cite{Bunk:2004br}. 
To construct a lattice fermion with a more physical interpretation, begin by
noting that the operator is $\gamma_5$-hermitian, namely 
\begin{equation}
  Q^\dagger = \gamma_5 Q \gamma_5,
\end{equation}
so that 
\begin{equation}
\det Q^\dagger = \det Q,
\end{equation}
and 
\begin{equation}
\sqrt{\det Q^\dagger Q} = \det Q.
\end{equation}

The product $\Box = Q^\dagger Q$ is diagonal in the spinor index, and has the 
form
\begin{equation}
\Box = \sum_\mu \Box_\mu.
\end{equation}
The lattice interaction $\Box$ thus resembles $2^d$ distinct copies of the 
simple discretisation of the continuum Klein-Gordon operator, $-\nabla^2+m^2$ 
on each of the $2^d$ lattices with spacing $b=2a$.

\section{An equivalent local Dirac operator \label{sec:dirac}}

In order to define a theory with a single flavour of fermion, the standard
method is to consider the appropriate fractional powers of the fermion
determinant. A single flavour of staggered fermion would then be represented by 
$\det Q^{1/n_t}$. It is important to recognise the significant theoretical 
difficulty with this prescription: the fractional power of the
determinant can no longer be written directly as a path integral over Grassmann
fields coupled through a local operator ($Q^{1/n_t}$ is non-local) and hence 
all the standard quantum field theory mechanisms for generating correlation 
functions by adding sources to the path integral no longer follow. 
Locality ensures that interacting theories are in the same universality
class of the continuum field theory.  In order to 
define a sensible lattice quantum field theory, an operator with the property 
\begin{equation}
  \det D = \det Q^{1/n_t} \label{eqn:equiv},
\end{equation}
is required, where $D$ defines local interactions \cite{DeGrand:2003xu}. With 
this property a path integral representation can be made, namely
\begin{equation}
  \det Q^{1/n_t} = \int\!\!{\cal D}\bar{\psi}{\cal D}\psi \; \exp\left\{-\bar{\psi} D \psi\right\}.
\end{equation}
Given this form, correlation functions of the theory can then be constructed by
adding source terms and following the standard construction. In this section,
an operator $D$ obeying Eq. \ref{eqn:equiv} is defined numerically for the 
free staggered fermion theory.

One observation in beginning the construction of $D$ is helpful: the operator 
will not obey the staggered-fermion Dirac algebra (which scatters the spin and 
taste components over the corners of the unit hypercube) since a counting of 
degrees-of-freedom suggests there are certain to be too many flavours. 
Instead, there must be sites on the lattice with no quark field. To construct 
a fermion field with the correct number of degrees of freedom in a 
hypercube requires there to be $2^{d/2}$ components, rather than the $2^d$ 
components of the staggered field. To begin construction assume there is a 
single Dirac spinor at one site per hypercube, with no degrees of freedom on 
the other sites of the cell. 

The equivalence property of Eq. \ref{eqn:equiv} is sufficient to define the
path integral, but can be trivially satisfied for the free theory: any
non-singular matrix can be made to obey this constraint after a rescaling. For
the free case a more stringent definition of equivalence must be made, namely
that the energy-momentum dispersion relation for fermions in the two theories 
be related. This will be satisfied if the operator itself squares to the free
Klein-Gordon operator on the blocked lattice, {\it i.e.} if (for massless 
fermions)
\begin{equation}
 D^\dagger D = -\square.
   \label{eqn:strong-equiv}
\end{equation}
In this work, this property will be denoted ``strong'' equivalence, while the
condition of Eq. \ref{eqn:equiv}, which is trivial for free fermions but
non-trivial in the interacting theory is denoted ``weak'' equivalence.

The properties expected from a well defined lattice Dirac operator
$D$ are locality, the correct continuum limit for momenta below the cutoff,
$\pi/a$ and invertibility at all non-zero momenta. Once these
properties are satisfied the Nielsen-Ninomiya theorem~\cite{Nielsen:1981hk} 
excludes the possibility of having invariance under continuous chiral
transformations. This last issue will be dealt with in the next section.
Here, before beginning to describe in detail our proposal of a Dirac
operator it is helpful to state what locality means.
An action density which has nearest-neighbour interactions or
interactions that are identically zero beyond a few lattice units is
certainly local, but no physical principle requires this
extreme case~\cite{Hernandez:1998et}.
On the lattice an action is termed local if its couplings have
exponentially decaying tails at large distances. This property is
ensured if $D(p)$ is an analytic periodic function of the momenta
$p_\mu$ with period $2\pi/a$.\\

The following ansatz for a solution to the ``strong'' equivalence constraint 
for the blocked lattice with spacing $b$ is made:
\begin{equation}\label{eqn:dirac}
D =  \gamma_{\mu} {p}_{\mu} -{q},
\end{equation}
with ${p}_{\mu}$ and ${q}$ such that $D$ obeys 
Eq. \ref{eqn:strong-equiv}, so 
\begin{equation}\label{op1}
{p}_{\mu}{p}_{\mu} - {q}^{2} = \square.
\end{equation}

A numerical prescription for constructing an effective representation of the
Dirac operator for massless fermions is used.  To begin, a sequence of 
``ultra-local'' operators of finite, increasing range is defined.  
The finite range operator can be described with a number of coefficients
weighting each distinct hopping term. The hopping terms are taken from $A_r$,
the set of all vectors ${\mathbf a}$ whose range is less than ${\mathbf r}$. 
The ``taxi-driver'' metric is used to define the range of a vector 
${\mathbf a}$ with components $a_i$, so
\begin{equation}
r({\mathbf a}) = \sum_i |a_i|.
\end{equation}
In two dimensions then, 
\begin{eqnarray}
A_0 &=& \left\{ (0,0)\right\}, \nonumber \\
A_1 &=& \left\{ (0,0), (1,0), (-1,0), (0,1), (0,-1) \right\}, \dots
\end{eqnarray}
A general ansatz for both $p^\mu$ and $q$, connecting fields at
sites ${\mathbf x}$ and ${\mathbf y}$  is then
\begin{equation}
   p^\mu_{{\mathbf x},{\mathbf y}} 
     = \sum_{{\mathbf a} \in A_r} \omega^\mu_p({\mathbf a}) 
      \delta_{ {\mathbf x}+{\mathbf a},{\mathbf y} },
\label{eqn:pmu}
\end{equation}
and
\begin{equation}
   q_{{\mathbf x},{\mathbf y}} = 
      \sum_{{\mathbf a} \in A_r} \omega_q({\mathbf a}) 
      \delta_{ {\mathbf x}+{\mathbf a},{\mathbf y} }.
\label{eqn:q}
\end{equation}
The coefficients are constrained so that the required symmetries of each
operator are preserved to ensure the action is a scalar. This implies that the
coefficients $\omega_q$ form a trivial representation of the lattice rotation
group, while $\omega^\mu_p$ form a fundamental representation. In two
dimensions (where the relevant rotation groups is $C_{4\nu}$) the required 
irreducible representations are $A_1$ for $q$ and $E$ for $p$. This in turn
implies the relations
\begin{widetext}
\begin{eqnarray}
      \omega_q( a_1, a_2) = 
      \omega_q( a_1,-a_2) = 
      \omega_q(-a_1, a_2) = 
      \omega_q(-a_1,-a_2) = \nonumber \\
      \omega_q( a_2, a_1) = 
      \omega_q( a_2,-a_1) = 
      \omega_q(-a_2, a_1) = 
      \omega_q(-a_2,-a_1),
\end{eqnarray}
and
\begin{eqnarray}
      \omega^1_p( a_1, a_2) = 
      \omega^1_p( a_1,-a_2) = 
     -\omega^1_p(-a_1, a_2) = 
     -\omega^1_p(-a_1,-a_2) = \nonumber \\
      \omega^2_p( a_2, a_1) = 
     -\omega^2_p( a_2,-a_1) = 
      \omega^2_p(-a_2, a_1) = 
     -\omega^2_p(-a_2,-a_1).
\end{eqnarray}
\end{widetext}
One further constraint is added to improve the representation of low-momentum
states. The coefficient $\omega_q(0,0)$ is chosen such that the operator
$q$ vanishes on a zero-momentum plane-wave. 
The number of free parameters in the operators $p$ and $q$ in two and four
dimensions is given for a few low ranges in Table \ref{tab:nops}.
\begin{table}[h]
\begin{tabular}{ccccc}
\hline
      & 
\multicolumn{2}{c}{\ \hspace{2em}d=2\hspace{2em}\ } &
\multicolumn{2}{c}{\ \hspace{2em}d=4\hspace{2em}\ }\\
\cline{2-5} 
\hspace{1em}Range\hspace{1em}\ &  $p^\mu$ & $q$          & $p^\mu$ & $q$          \\
\hline
  1   &  1       &   1          & 1       &   1          \\
  2   &  3       &   2          & 3       &   3          \\
  3   &  6       &   4          & 7       &   6          \\
  4   &  10      &   6          & 14      &  11          \\
  5   &  15      &   9          & 25      &  17          \\
 10   &  55      &  30          & 189     &  93          \\
\hline
\end{tabular}
\caption{The number of free parameters in the finite-range operators in two and
four dimensional lattice actions\label{tab:nops}}.
\end{table}

A sequence of lattice Dirac operators, $D_1, D_2, \dots$ with increasing range 
is then considered.  Each operator in the sequence is chosen to minimise 
$\mu_r^2$, a positive-definite measure of the difference between the two sides 
of Eq. \ref{eqn:strong-equiv}, namely
\begin{equation}
   \mu_r^2 = \frac{1}{4d^2N_s} \mbox{Tr }(X_r^2),
    \label{eqn:mu}
\end{equation}
with $N_s$ the number of sites on the blocked lattice and
\begin{equation}
   X_r = D_r^\dagger D_r + \square.
\end{equation}

Note there are certainly an infinite number of actions obeying the 
equivalence principle of Eq. \ref{eqn:strong-equiv}. Most of these will be 
non-local but there could well be more than one local action.
The following hypothesis is made. If a local action obeying ``strong'' 
equivalence exists, then the measure $\mu_r$ should fall exponentially, and the 
operator $D_r$ should have exponentially falling coefficients inside $A_r$. 
$D_r$ is the best ultra-local approximation to the solution of the equivalence 
condition of Eq. \ref{eqn:strong-equiv}. The coefficients of the action a long 
way from the boundary of the operator should also converge as the range is
increased. 

The sequence of ultra-local actions $D_r$ was computed numerically by finding
the minimum of $\mu_r$. The calculations were performed for massless fermions.
A short check demonstrated the localisation properties were better for massive
fermions. 

\subsection{Results}
A multi-dimensional Newton-Raphson solver was used, since both the slope and
Hessian of $\mu_r$ can be computed easily. The GNU multiple precision library 
(GMP) was used \cite{GMP} when numerical precision was required beyond 64-bit 
native arithmetic.  Some checks were made to test if the minimum in $\mu_r$ was 
a global one. A range of different starting values of the action
parameters were used to seed the Newton-Raphson search and a simulated
annealing algorithm was run to search for a minimum at short ranges. A number
of local minima were found in many cases making it difficult to determine if
the global minima was reached. This issue is discussed later.

\subsubsection{Two dimensions}

\begin{figure}[t]
\includegraphics[width=8cm]{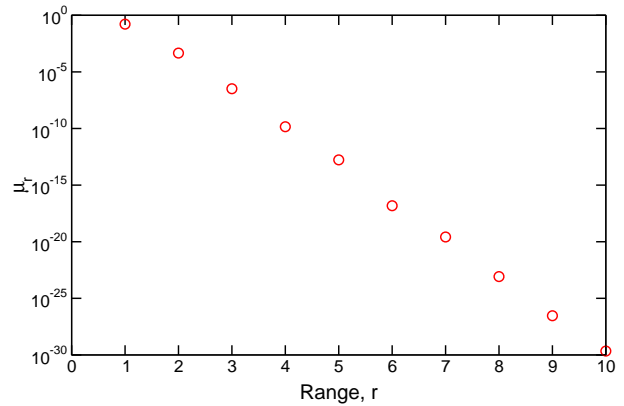}
\caption{The error function, $\mu_r$ for lattice actions of range up to $10b$.
\label{fig:mu-Q2d}}
\end{figure}
Fig. \ref{fig:mu-Q2d} shows the dependence of $\mu_r$ for the optimal action as a
function of the finite range of the action, $r$. A clear signal for exponential
fall-off is displayed: $\mu_r$, the discrepancy between $D_r^\dagger D_r$ and
$-\square\;$ falls by thirty decades as the action range is increased from $b$ 
to $10b$. 
\begin{figure}[t]
\includegraphics[width=8cm]{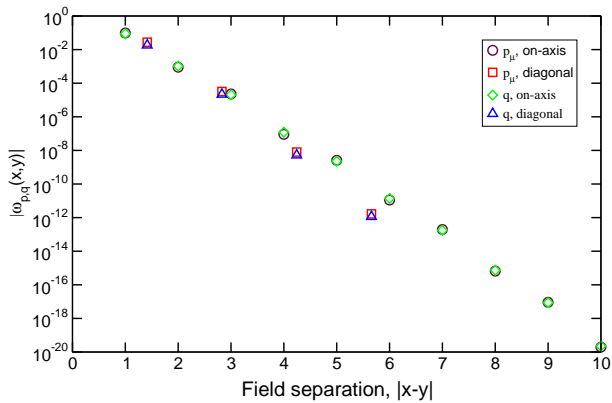}
\caption{The coefficients in $q$ and $p_\mu$, the composite operators in the 
action $D_{10}$, as a function of the separation between the fields in the 
bilinear. The field separations are given using the usual 2-norm distance.
\label{fig:omega-Q2d}}
\end{figure}
The coefficients in $q$ and $p_\mu$ of the action $D_{10}$ are presented in 
Fig. \ref{fig:omega-Q2d}. The on-lattice-axis and diagonal terms are presented. For
the operator $p_\mu$, on-axis refers to the terms in $p_1$ with off-set vector
$(a,0)$ and those in $p_2$ with off-set $(0,a)$. Note that by symmetry, terms
in $p_1$ with off-set $(0,a)$ vanish identically.  An exponential fall-off 
over eighteen decades is observed, providing solid evidence for the existence 
of a local operator. At 64-bit machine precision, terms with ranges beyond 
about $6b$ would have uncomputably small contributions to the action of a
plane wave state. Notice the terms in the two operators $p_\mu$ and $q$ are 
very similar in magnitude and have similar localisation range. 
\begin{figure}[t]
\includegraphics[width=8cm]{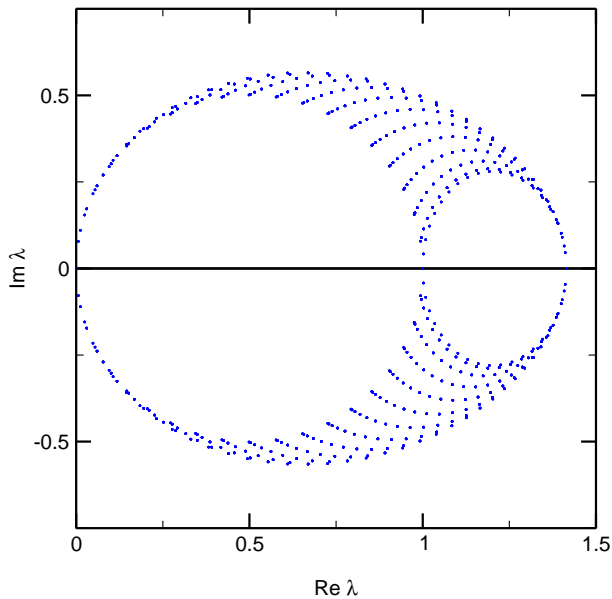}
\caption{The eigenvalues of the approximately equivalent Dirac operator, 
$D_{10}$. The eigenvalues are scaled so the first doubler, with momentum 
$(\pi,0)$ has $\lambda=1$.
\label{fig:evals-Q2d}}
\end{figure}
Fig. \ref{fig:evals-Q2d} shows the eigenvalue spectrum of the operator
$D_{10}$. The eigenvalues are purely real when all components of the operator 
$p_\mu$ vanish, {\it ie.} at the doubling points. The real parts of the
eigenvalues for the doublers are close to 0 (for the propagating mode), 1 and
$\sqrt{2}$. 

\subsubsection{Four dimensions}

A similar construction was followed for the four-dimensional action. A sequence
of operators on the four-dimensional lattice with spacing $b=2a$ was
determined. The operators in the action again took the general form of Eqns.
\ref{eqn:pmu} and \ref{eqn:q}, with $q$ transforming trivially under the
four-dimensional rotation group and $p_\mu$ transforming under the fundamental 
representation.
The computational cost of the four-dimensional calculation means that only
actions up to range $r=5$ have been constructed. 
\begin{figure}[t]
\includegraphics[width=8cm]{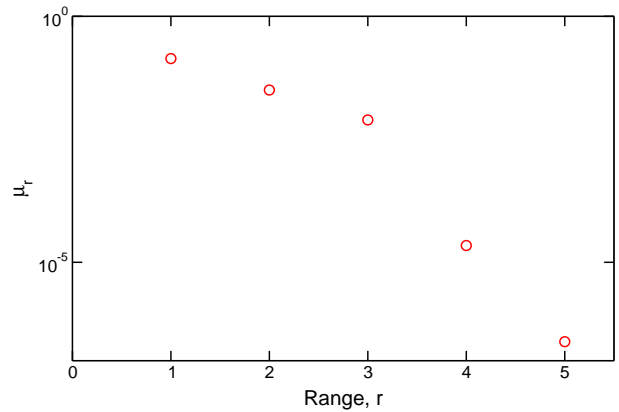}
\caption{The error function, $\mu_r$ for four-dimensional lattice actions of
range up to $5b$.
\label{fig:mu-Q4d}}
\end{figure}
\begin{figure}[t]
\includegraphics[width=8cm]{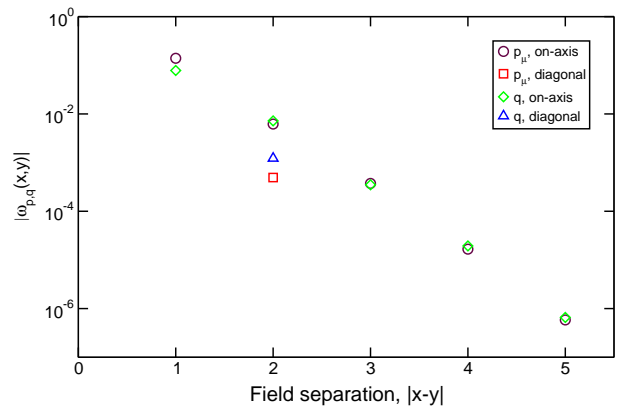}
\caption{The coefficients in $q$ and $p_\mu$ in the four-dimensional action
$D_5$ as a function of the separation between the fields in the bilinear
\label{fig:omega-Q4d}}.
\end{figure}
Fig. \ref{fig:mu-Q4d} shows the fall-off of $\mu_r$ in the sequence of actions.
A rapid decay of $\mu_r$ as $r$ is increased is again observed. Notice also the
fall-off accelerates at range $r=4$, once the action increases beyond the 
bounds of the unit hypercube on the spacing-$b$ lattice.

In Fig. \ref{fig:omega-Q4d} the magnitude of the coefficients on the axis and 
diagonals of the operator $D_5$ are displayed. A similar pattern to the
two-dimensional case is seen, with a five-decade decay over five lattice hops
being observed. The data suggest a local operator exists in four dimensions as
well as two. 

\section{The Ginsparg-Wilson relation \label{sec:gw}}

The presence of the operator ${q}$ in the definition of
Eq. \ref{eqn:dirac}, means $D$ does not anticommute with $\gamma_5$, which 
would guarantee that the fermionic action is invariant under continuous chiral 
transformations.  This is expected from the Nielsen-Ninomiya theorem.
As is now well known, the theorem can be bypassed if one does not insist
that the chiral transformations assume their canonical form on the
lattice~\cite{Ginsparg:1981bj, Luscher:1998pq}. In particular it was shown that 
the Ginsparg-Wilson relation,
\begin{equation}\label{eqn:GW}
\{ \gamma_5 , D \} = 2 D \gamma_5 R D,
\end{equation}
implies an exact symmetry of the fermionic action 
which may be regarded as a lattice form of an infinitesimal chiral rotation.

For a Dirac operator obeying Eq. \ref{eqn:dirac}, three properties follow:
\begin{eqnarray}
D^{\dagger}   &=& \gamma_5 D \gamma_5,\\
D^{\dagger} D &=& -\square \ \mathcal{I},\\
D^{-1}        &=& \frac{\gamma_\mu {p}_\mu + {q}}
                     {\Box}.
\end{eqnarray}
The propagator $D^{-1}$ then satisfies the following relation
\begin{equation}
\{ \gamma_5 , D^{-1} \} = 2 R \gamma_5,
\end{equation}
with 
\begin{equation}
R = \frac{{q}}{\square}.
  \label{eqn:Rdef}
\end{equation}

The construction of Sec. \ref{sec:dirac} does not ensure the operator $R$ is 
local, and so there is no apparent lattice chiral symmetry. Eqn. 
\ref{eqn:Rdef} does suggest the definition of an alternative sequence of 
actions, $D^{(GW)}_r$ which might lead to a Dirac operator obeying a 
(generalised) Ginsparg-Wilson relation. Consider an operator of range $r$ 
with the form 
\begin{equation}
  D^{(GW)}_r = \gamma_\mu p^{\mu}_r + \square R_{r-1},
    \label{eq:D-GW}
\end{equation}
where $R_{r-1}$ is a local operator with finite range $r-1$. Note this implies 
the 
scalar operator $q_r = \square R_{r-1}$ has range $r$ as before. If the limit
of the sequence $D^{(GW)}_r$ is a solution to Eqn. \ref{eqn:dirac} and $p^{\mu}$
and $R$ remain local, then a local operator, equivalent to the staggered
fermion and obeying a generalised Ginsparg-Wilson relation will be constructed.

A simple consequence of Eq. \ref{eqn:GW} is then 
that chiral symmetry is partly preserved, in particular the 
lagrangian $L =  \overline{\psi} D \psi $ is invariant under the
local symmetry transformation:
\begin{equation}
\delta \psi = \gamma_5 \left( 1 - \frac{1}{2}
R D \right) \psi,
\end{equation}
\begin{equation}
\delta \overline{\psi} = \overline \psi \left( 1 - \frac{1}{2}
D R \right) \gamma_5.
\end{equation}

\subsection{results}

The operator sequence minimising $\mu_r$ of Eq. \ref{eqn:mu} was computed for
the two-dimensional lattice.  As before, the Newton solver was used for
minimisation with the constraint of Eqn. \ref{eq:D-GW}. The problem of finding 
multiple local minima was observed and was more extreme than the initial 
construction of Sec. \ref{sec:dirac}. 
\begin{figure}[t]
\includegraphics[width=8cm]{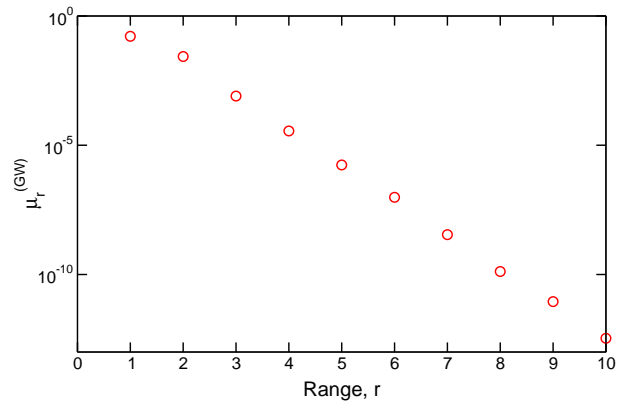}
\caption{The error function, $\mu_r$ for two-dimensional lattice operators
constructed with the Ginsparg-Wilson constraint of Eqn. 
\protect{\ref{eq:D-GW}}.
\label{fig:mu-R2d}}
\end{figure}
\begin{figure}[t]
\includegraphics[width=8cm]{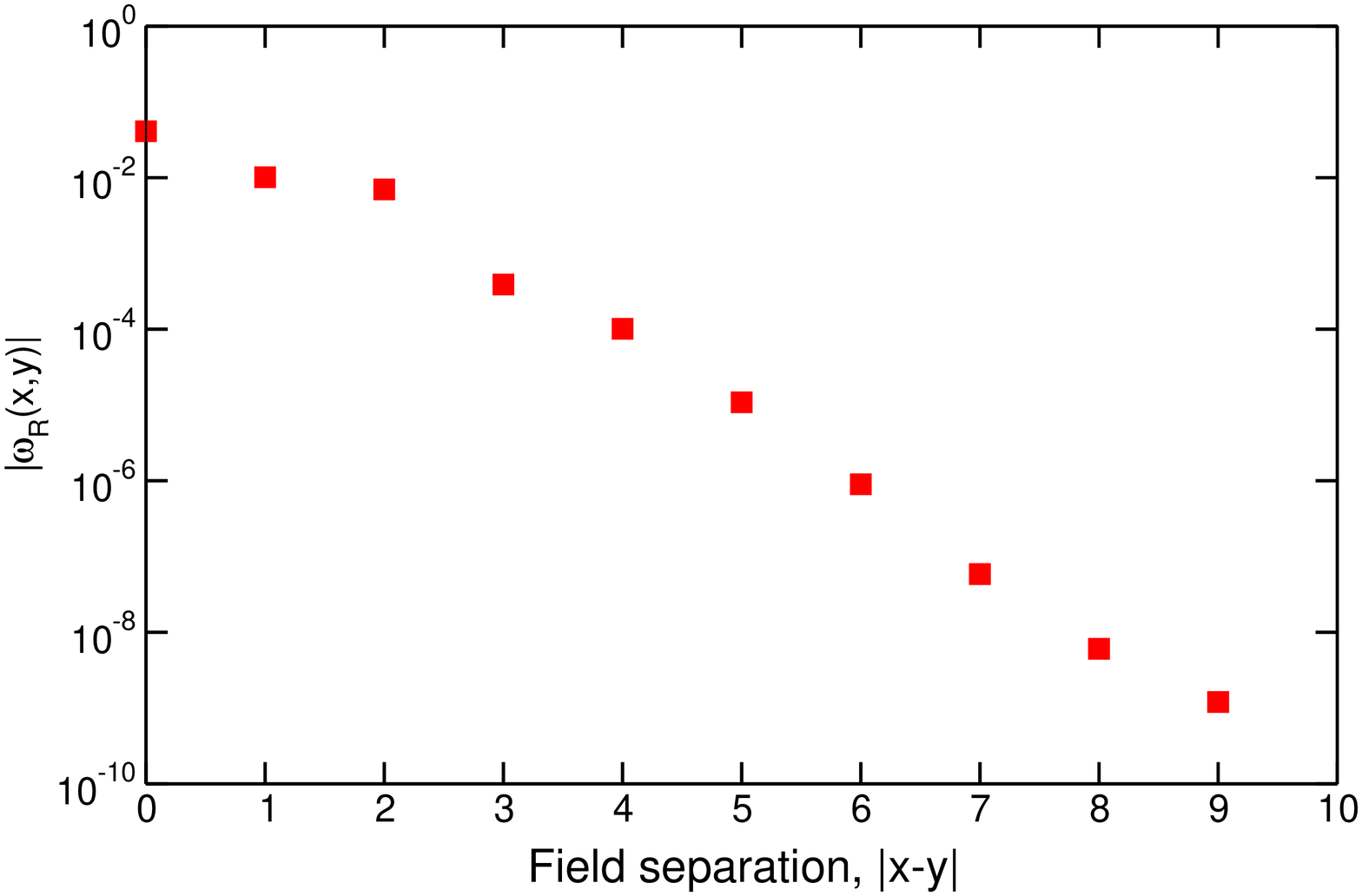}
\caption{The coefficients in $q$ and $p_\mu$ in the two-dimensional action
$D^{(GW)}_{10}$ as a function of the separation between the fields in the 
bilinear.
\label{fig:omega-R2d}}
\end{figure}
\begin{figure}[t]
\includegraphics[width=8cm]{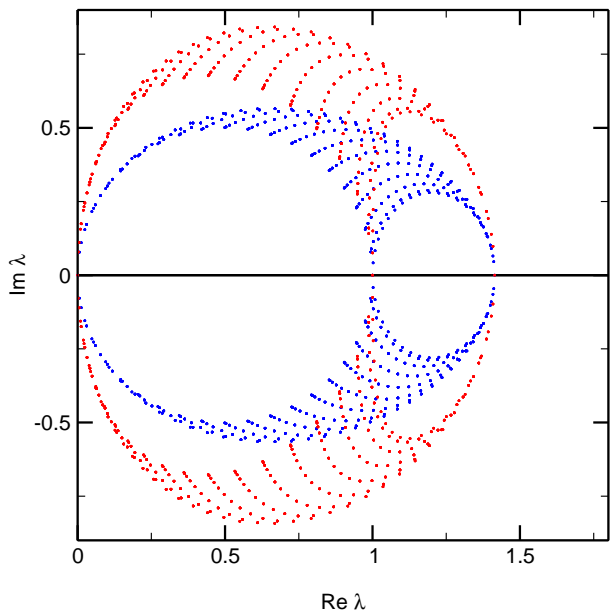}
\caption{The eigenvalues of the two operators, $D_{10}$ (blue) and 
$D^{(GW)}_{10}$ (red). The eigenvalues are scaled such that $\lambda=1$ for 
momentum $(\pi,0)$.
\label{fig:evals-R2d}}
\end{figure}
The dependence of $\mu_r$ for the operator obeying the Ginsparg-Wilson
constraint is shown in Fig. \ref{fig:mu-R2d}. Exponential decay of the
discrepancy measurement, $\mu_r$ as the range $r$ is increased is seen again
this time over twelve orders-of-magnitude. 
Fig. \ref{fig:omega-R2d} shows the fall-off of the symmetry-breaking kernel for
the optimised action $D^{(GW)}_{10}$. An exponential fall-off of eight decades
is observed. The terms in the operator $q$ are thus exponentially localised.
The derivative operator, $p_\mu$ also has local coefficients. 
The eigenvalues of the operator $D^{(GW)}_{10}$ are shown in Fig.
\ref{fig:evals-R2d}, along with those for the unconstrained two-dimensional
action construction of Sec. \ref{sec:dirac}. 

\section{Discussion \label{sec:discuss}}

A numerical construction of ultra-local, approximate actions can never prove
the existence of a fermion with a well defined local action (unless the action
is itself ultra-local), but the calculations in this paper do present
strong evidence for the existence of an equivalent, local theory to the
one-flavour free staggered fermion. In two dimensions, the mis-match between
the dispersion spectrum of the ultra-local theory and the staggered fermion can
be made as small as $10^{-30}$ with an action of range $10b$. At this range,
terms in the action are as small as $10^{-18}$. The construction for 
four-dimensional theories is more difficult, but evidence for locality is seen 
here too. 
The numerical construction of the action was performed for massless fermions.
Some short tests with massive fermions suggest the localisation properties of
these actions are better still; the massless fermion represents the hardest
case to reproduce. 

The numerical search for a global minima of Eqn. \ref{eqn:mu}, a non-linear
function of the action parameters, $\omega_p$ and $\omega_q$ is made difficult
by the presence of local minima. It is difficult to find convincing evidence
that the Newton-Raphson solver has found the global minimum for large actions,
although searches using different starting points often converged to a common
fixed point. An empirical observation is important; the minima with smaller
values of $\mu_r$ tend to have better localisation properties (their
coefficients fall faster). This implies if the searches have not found the
global minima, these will represent better actions than those already
uncovered, improving the construction rather than spoiling it. 

For the constrained construction to build an operator obeying the (generalised)
Ginsparg-Wilson relation, the localisation ranges were about twice that of the
unconstrained construction and good evidence for exponentially local actions 
is seen. The problem of finding global minima seems to be exacerbated. 
Solutions to the standard GW relation (with $R=I$) are now well known
\cite{Neuberger:1997fp,Hasenfratz:2000xz}.
It is important to recognise a solution with $R=I$ is impossible for the 
equivalent theory; this can be seen easily by considering the doubler momenta,
$(\pi,0)$ and $(\pi,\pi)$. For these two points, the eigenvalues of $D$ must be
$1$ and $\sqrt{2}$ respectively, while $R=I$ would demand they were both unity.

The evidence in this paper (and in Ref. \cite{David}) put staggered-fermion
simulations on a more robust footing, but there remain many unanswered
questions. For staggered fermion simulations to be correct descriptions of
quantum field theory, one must demonstrate two postulates; firstly that a local
path-integral representation of the fractional power of the staggered
determinant exists and secondly the validity of calculations 
performed by assuming the propagator of this theory is related to the 
inverse of the 
full staggered matrix, $Q^{-1}$. The work in this paper hints at the right
question to address the first issue, but does not address the second point. 
In using $Q^{-1}$ as the fermion propagator a clear problem arises. In four
dimensions, too many pion operators can be constructed for example. The
residual symmetry of the staggered matrix ensures these states lie in
mass-degenerate multiplets \cite{Golterman:1986jf, Lee:1999zx, Orginos:1999cr}, 
but ``taste-breaking'' interactions split their masses. These splittings 
vanish in the continuum limit. Recent work on the low-energy eigenvalue 
spectrum is beginning to resolve this issue \cite{Durr:2003xs,Durr:2004as,
Follana:2004wp}. 

Future work offers an optimistic possible outcome; if an effective operator can
be constructed for QCD, it seems this operator might obey a
Ginsparg-Wilson relation and the lattice physicist will have the best of both
worlds: cheap dynamical-configuration generation algorithms using the staggered
formulation with a theoretically well defined action (possibly with an exact
GW chiral symmetry) to compute propagators. This would not be a ``mixed action''
simulation, where different valence and sea quark actions are employed. The 
first obstacle to extending the construction of Sec. \ref{sec:dirac} to 
incorporate gauge interactions is that the operator $Q^\dagger Q$ is not 
proportional to $(I \otimes I)$ in Dirac-taste indices and so does not 
decompose directly into $n_t$ decoupled parts. This is the ``strong'' 
definition of equivalence required for the free theory, but a re-definition of 
$\mu_r$ to measure violations in ``weak'' equivalence can be made. This is 
under investigation and few conclusions about the success of this programme can 
be drawn. A number of difficult questions arise immediately, since the two 
theories would have different apparent symmetries. 

\section{Conclusions \label{sec:conclude}}

In this paper, a local lattice Dirac operator whose determinant is identical 
to the free staggered-fermion determinant, and whose energy-momentum dispersion
relations are identical
(although with different degeneracies) is described as the end-point of a 
sequence of actions of increasing, finite range. The first few ultra-local 
actions in the sequence are constructed numerically and convergence of 
the sequence is demonstrated in both two and four dimensions.

The spectrum of the 
operator is free from doublers and its low-energy dynamics correctly describes 
the propagation of free fermions up to corrections of ${\cal O}(a^2 p^2)$, a
property it inherits from its staggered parent. The operator acts on a full 
Dirac spinor situated only on the sites of a blocked lattice with spacing 
$b=2a$. 

A constraint is added to the construction to define an operator that obeys a 
generalised Ginsparg-Wilson relation. In this action, the chiral symmetry 
breaking in the propagator is described by a local operator, diagonal in the 
spin index.  This implies the existence of a fermion whose path integral is the
same as that of one staggered flavour and with an exact chiral symmetry on the 
lattice.

\section*{Acknowledgements}

We are grateful to David Adams for helpful discussions about the contents of 
his manuscript \cite{David} prior to publication. We would also like to thank 
Jonathan Bennett for a discussion about operator locality. This work was 
supported by Enterprise-Ireland and the Irish Higher Education Authority 
Programme for Research in Third-Level Institutes (PRTLI).

\end{document}